\documentclass[%
jcp,%
 amsmath,amssymb,
preprint,
onecolumn,
]{revtex4-1}

\usepackage{graphicx}
\usepackage{dcolumn}
\usepackage{bm}

\begin{document}

\preprint{Preprint}

\title{Influence of solvent polarization and non-uniform ion size on electrostatic properties between charged surfaces in an electrolyte solution}

\author{Jun-Sik Sin}

 \email{js.sin@ryongnamsan.edu.kp}

  \affiliation{Department of Physics, Kim Il Sung University, Taesong District, Pyongyang, Democratic People's Republic of Korea}


\begin{abstract}
In this paper, we study electrostatic properties between two similar or oppositely charged surfaces immersed in an electrolyte solution by using mean-field approach accounting for solvent polarization and non-uniform size effect. Applying a free energy formalism accounting for unequal ion sizes and orientational ordering of water dipoles, we derive coupled and self-consistent equations to calculate electrostatic properties between charged surfaces. 
Electrostatic properties for similarly charged surfaces depend on counterion size but not coion size. Moreover, electrostatic potential and osmotic pressure between similarly charged surfaces are found to be increased with increasing counterion size. On the other hand, the corresponding ones between oppositely charged surfaces are related to both sizes of positive and negative ions. For oppositely charged surfaces, the electrostatic potential, number density of solvent molecules and relative permittivity of an electrolyte having unequal ion sizes are not symmetric about the centerline between the charged surfaces.
For either case, the consideration of solvent polarization results in an decrease in the electrostatic potential and the osmotic pressure compared to the case without the effect.
\end{abstract}
\maketitle

\section{\label{sec:level1}Introduction}
The study of electrostatic properties between charged surfaces in an electolyte is of great significance in material science and biology.
It is well known that the interaction between two charged surfaces is attributed to two kinds of physical natures. (i.e., Van der Waals and Electrostatic interaction).\cite%
{Israel_book_1985, Lyklema_book_2005, Hunter_book_1981, Hansen_ARPC_2000, Kampf_PRL_2009}

Although classical Poisson-Boltzmann (PB) theory has been a fundamental tool describing electric double layer electrostatic potential and osmotic-pressure between two charged surfaces for a long time \cite%
{Gouy_JPhysF_1910, Chapman_PhilosMag_1913}, many researchers have devoted a great deal of effort to amending the PB theory which cannot be applicable to the case being short the distance between two charged surfaces. To  properly describe the electric double layer properties, there exist attempts to develop mean-field theories describing finite sizes of ions and water molecules and/or water polarization\cite%
{Bikerman_PhilosMag_1942, Wicke_ZEC_1952, Iglic_JPhysF_1996,Andelman_PRL_1997}.
In the last decade, the consideration of non-uniform ion sizes has attracted massive attention \cite%
{Chu_BJ_2007, Kornyshev_JPCB_2007, Biesheuvel_JCIS_2007, Li_PRE_2011, Li_PRE_2012, Boschitsch_JCC_2012, Siber_PRE_2013, Gongadze_EA_2015, Mohajernia_EA_2017}.
In particular, the authors of \cite%
{Iglic_Bioelechem_2010, Iglic_Bioelechem_2012} simultaneously accounted for orientational ordering of water dipoles and finite sizes of ions and water molecules.
 
Recently,  it was demonstrated in the papers \cite%
{Sin_EA_2015, Gongadze_EA_2015, Sin_PCCP_2016_1, Sin_EA_2016, Sin_PCCP_2016_2, Sin_CSA_2017,  Mohajernia_EA_2017} that the simultaneous considerations of unequal ion size and solvent polarization are very important for describing electrostatic properties of an electrolyte near a single charged surface.

On the other hand, the electrostatic properties between charged surfaces were studied by using mean-field theories taking into account finite ion size, solvent polarization and composition of solvent mixture.
Using a mean-field theory accounting only for ion size effect, the authors of \cite%
{Das_PRE_2011} demonstrated the valuable fact that nontrivial interactions between ion size effect and electric double layer overlap phenomena may augment the effective extent of electric double layer overlap in narrow fluidic confinements.
Considering ions as point-like charge, the authors of \cite%
{Das_PRE_2012} demonstrated the effect of solvent polarization on the electric double layer electrostatic potential distribution and the effective EDL thickness in narrow nanofluidic confinements.
Although the authors of \cite%
{Abrashkin_PRL_2007} addressed osmotic pressure between charged surfaces by accounting for dipole moment of water molecules and ion size, some of their results were contrary to common sense such as behavior of spatial distribution of permittivity. \cite%
{Teschke_CPL_2000, Teschke_JMCA_2001}  Namely, in \cite%
{Abrashkin_PRL_2007}, the increase of relative permittivity near the charged surface was predicted for point-like ions. This is a consequence of predicted accumulation of water dipoles near the charged surface due to an assumed Boltzmann distribution for water molecules, which prevails over the saturation effect in polarizability as shown in \cite%
{Gongadze_GPB_2011}.
The authors of \cite%
{Andelman_JPCB_2009} derived a general expression for osmotic pressure for the case when the free energy of the system does not depend explicitly on the coordinate.
Even though studies \cite
{Urbanija_JCP_2008, Andelman_PRE_2013, Bohinc_SM_2016, Andelman_PRE_2016} have been investigated electric double layer forces between charged surfaces, they did not considered the effect of solvent polarization. 

Recently, the authors of \cite%
{Das_JCP_2013, Gongadze_EA_2014} presented a more satisfactory answer on the osmotic pressure by using Langevin-Bikerman.
Their model well represents electrostatic properties by simultaneous considerations of size effects and polarization of water molecules, barring the difference in size between positive and negative ions. 

Although the authors of \cite%
{Gongadze_EA_2015, Mohajernia_EA_2017} described differential capacitance and permeation through charged nanotube membranes by accounting for both solvent polarization and disparity of ion sizes, there does not exist study to describe osmotic pressure between charged surfaces with consideration of the effects.
On the other hand,  Monte Carlo simulation \cite%
{Zelko_JCP_2010, Guerrero_JCP_2011, Wang_JCP_2012}was extensively used to correct the theoretical predictions of classical PB method. In fact, discreteness-of-charge and the image effects, considered by using Monte Carlo method, are of important roles in the compact part of the electric double layer and may substantially affect the zeta potential of the surface. In the present paper, these effects are beyond our scope and we provide results only for constant zeta potential because compared to such effects, ion size asymmetry and solvent polarization crucially affect electrostatic interaction between charged surfaces for high electrolyte concentration.

In this paper, we study the effect of the difference in size between positive and negative ions as well as solvent polarization on electrostatic potential, number density of ions and water molecules, permittivity and osmotic pressure by using a mean field approach. \cite%
{Iglic_Bioelechem_2010, Iglic_Bioelechem_2012, Sin_EA_2016, Gongadze_EA_2014, Gongadze_EA_2015} 
The first result is that evaluating electrostatic properties between oppositely charged surfaces requires considering difference in size between positive and negative ions, while these between similarly charged surfaces can depend only on counterion size. Next,  it is shown that  for a constant surface potential, solvent polarization diminishes ion size effects on electrostatic properties between similar and oppositely charged surfaces. 
Finally, we emphasize that our method can consistently explain the experimental results of interaction force between similar or oppositely charged surfaces, by correcting the large under-prediction made by the corresponding PB model or the over-prediction made by considering only the ion size. 

\section{Theory}
We consider two parallel plates (similar or oppositely charged) separated by a distance $H$ in an electrolyte. The transverse direction is noticed by x; the left plate is placed at $x=-H/2$ and the right plate $x=-H/2$.  The resulting electrostatic properties at the interfaces between the plates and the electrolyte solution should be addressed by setting the free energy of total thermodynamic system as follows    
\begin{eqnarray}
F = \int d {\bf{r}}\left[ { - \frac{{\varepsilon _0 E}}{2}^2  + e_0 z\psi \left( {\bf{r}} \right)\left( {n_ +   - n_ -  } \right) + \left\langle {p_0 E\cos \omega \rho \left( \omega  \right)} \right\rangle _\omega   - \mu _ +  n_ +   - \mu _ -  n_ -   - \left\langle {\mu _\omega  \left( \omega  \right)\rho \left( \omega  \right)} \right\rangle _\omega   - Ts} \right].
\label{eq:1}.
\end{eqnarray}
While the local electrostatic potential is denoted by $\Psi \left( {\bf{r}} \right)$, the number density of different ionic species and the number density of water molecules  are expressed as $n_i \left( {\bf{r}} \right), i = +,-
$ and $n_w \left( {\bf{r}} \right) = \left\langle {\rho \left( {\omega ,{\bf{r}}} \right)} \right\rangle _\omega$, respectively.

Here $\left\langle {f\left( \omega  \right)} \right\rangle _\omega   = \int {f\left( \omega  \right)2\pi \sin \left( \omega  \right)d\omega }$  and $\omega$  stands for the angle between the dipole moment vector {\bf{p}} and the normal to the charged surface. ${\bf{p}}$  and ${\bf{E}}$ notice the dipole moment of water molecules and electric field strength, respectively, where $p_0  = \left| {\bf{p}} \right|$ and $E = \left| {\bf{E}} \right|$.

In Eq. (\ref{eq:1}), the first term describes the self energy of the electrostatic field, where  stands for the vacuum permittivity.  The second term means the electrostatic energy of the ions. It is noticeable that unlike the case in \cite%
{Sin_EA_2016}, the the third term i.e., electrostatic energy of water dipoles, is equal to one of \cite%
{Iglic_Bioelechem_2010, Gongadze_EA_2014} where the formula for osmotic pressure was derived.   
  Coupling the system to a bulk reservoir necessitates the next three terms, where  $\mu _+$ and $\mu _-$ mean the chemical potentials of positive and negative ions, respectively, and $\mu_w\left(\omega\right)$   corresponds to the chemical potential of water dipoles with orientational angle $\omega$ . $T$  and $s$ are the temperature and the entropy density, respectively. 
(Please refer \cite%
{Iglic_JPhysF_1996, Iglic_Bioelechem_2010, Iglic_Bioelechem_2012, Sin_EA_2016}.)
From this free energy functional, we derive the self-consistent equations determining electrostatic properties by performing minimization of the corresponding free energy describing the electric double layer and subsequently find the formula for the osmotic pressure.
The Lagrangian of the total system can be expressed so that the volume conservation is satisfied
\begin{eqnarray}
L = F - \int {\lambda \left( {\bf{r}} \right)} \left( {1 - n_ +  V_ +   - n_ -  V_ -   - n_w V_w } \right)d{\bf{r}},
\label{eq:2}
\end{eqnarray}
where $\lambda$ is a local Lagrange parameter. 
When the origin of the electric potential is located at $x = \infty$, $\psi \left( {x= \infty } \right)=0$. 
At the origin,   $n_i \left( {x = \infty } \right) = n_{ib}$ and $\lambda \left( {x = \infty } \right) = \lambda _b$, where $n_{ib}$ and $\lambda _b$ represent the bulk ionic concentration and the Lagrange parameter at $x = \infty$, respectively. 
The number densities of ions and water molecules can be obtained by applying the boundary conditions and by writing Euler-Lagrange equations of Eq. (\ref{eq:2}) in terms of the number densities of particles. 
\begin{subequations}
\label{eq:3}
\begin{equation}
n_ +   = {{n_{ + b} \exp \left( { - V_ +  h - e_0 z\beta \psi } \right)}}/{D},
\label{subeq:1}
\end{equation}
\begin{eqnarray}
n_ -   ={{n_{ - b} \exp \left( { - V_ -  h + e_0 z\beta \psi } \right)}}/{D},
\label{subeq:2}
\end{eqnarray}
\begin{eqnarray}
n_w  = {{n_{wb} \exp \left( { - V_w h} \right)\frac{{\sinh \left( {p_0 \beta E} \right)}}{{p_0 \beta E}}}}/{D},
\label{subeq:3}
\end{eqnarray}
\begin{eqnarray}
n_{ + b} \left( {e^{ - V_ +  h - \beta ze\phi }  - 1} \right) + n_{ - b} \left( {e^{ - V_ -  h + \beta ze\phi }  - 1} \right) + n_{wb} \left( {e^{ - V_w h} \frac{{\sinh (p_0 \beta E)}}{{p_0 \beta E}} - 1} \right) = 0,
\label{subeq:4}
\end{eqnarray}
\end{subequations}
where $D = n_{ + b} V_ +  \exp \left( { - V_ +  h - e_0 z\beta \psi } \right) + n_{ - b} V_ -  \exp \left( { - V_ -  h + e_0 z\beta \psi } \right) + n_w V_w \exp\left({-V_w h}\right)\frac{{\sinh \left( {p_0 \beta E} \right)}}{{p_0 \beta E}}$, $h = \lambda  - \lambda _b$ and $\left\langle {e^{ - p_0 E\beta \cos \left( \omega  \right)} } \right\rangle _\omega   = \frac{{2\pi \int_\pi ^0 {d\left( {\cos \omega } \right)e^{ - p_0 E\beta \cos \left( \omega  \right)} } }}{{4\pi }} = \frac{{\sinh \left( {p_0 E\beta } \right)}}{{p_0 E\beta }}$. \cite%
{Iglic_Bioelechem_2010, Iglic_Bioelechem_2012}

In addition to the above equations, we have the following equations for the chemical potentials of ions and water molecules
\begin{subequations}
\label{eq:5}
\begin{equation}
\mu _ +   = k_B T\ln \left( {n_{ + b} /N_b } \right) + V_ +  \lambda _b, 
\label{subeq:1}
\end{equation}
\begin{equation}
\mu _ -   = k_B T\ln \left( {n_{ - b} /N_b } \right) + V_ -  \lambda _b, 
\label{subseq:2}
\end{equation}
\begin{equation}
\mu _w\left(\omega_i\right)  = k_B T\ln \left( {\rho\left(\omega_i\right)\Delta\Omega /N_b } \right) + V_w \lambda _b,
\label{subseq:3}
\end{equation}
\end{subequations}
where $\rho\left(\omega_i\right)$  stands for the number density of water molecules with orientational angle $\omega_i$ and $n_{w}=\left\langle{\rho\left( \omega  \right)}\right\rangle_\omega$.

The Euler-Lagrangian equation for $\psi \left( r \right)$ yields the Poisson equation
\begin{equation}
\nabla \left( {\varepsilon _0 \varepsilon _r \nabla \psi } \right) =  - e_0 z\left( {n_ +   - n_ -  } \right),
\label{eq:6}
\end{equation}
where $\varepsilon _r  \equiv 1 + \frac{{\left| {\bf{P}} \right|}}{{\varepsilon _0 E}}$. 	
Due to the symmetry of the present study, $\bf{P}$, the polarization vector due to a total orientation of point-like water dipoles, is parallel to the normal to charged surfaces, as in \cite%
{Iglic_Bioelechem_2010, Das_JCP_2013, Gongadze_GPB_2011}
\begin{equation}
{\bf{P}}\left( x \right) = n_w \left(x \right)p_0 L\left( {p_0 E\beta } \right),
\label{eq:7}
\end{equation}
where $\beta  = 1/\left( {k_B T} \right)$, $p_0=4.8D$ and the function $L\left( u \right) = \coth \left( u \right) - 1/u$  means the Langevin function.

For now, let us derive a new formula for osmotic pressure accounting for unequal ion sizes and solvent polarization. 
In fact, the authors of \cite%
{Andelman_JPCB_2009} proved the fact that when the free energy density of the total system does not depend on the coordinate, osmotic pressure between two charged surfaces can be derived from the following expression
\begin{equation}
f - \left( {\partial f/\partial \psi '} \right)\psi ' = consant =  - P,
\label{eq:10}
\end{equation}
where $\psi'$ is the derivative of $\psi$ with respect to $x$ and the constant is the negative of the local pressure $P$ that is defined to be the sum of the osmotic pressure and the bulk pressure, i.e., 
\begin{equation}
P=P_{osm} + P_{bulk}.
\label{eq:13}
\end{equation}
The free energy density for the present study does not depend explicitly on the coordinate x, as can be seen in Eq.(\ref{eq:1}). 
Using Eqs. (\ref{eq:1}) and (\ref{eq:10}), we can therefore get
\begin{equation}
\left( {\partial f/\partial \psi '} \right)\psi ' =  - \varepsilon _0 \psi '^2  + \left\langle {\rho \left( \omega  \right)p_0 E\cos \omega } \right\rangle _\omega  
\label{eq:11}
\end{equation}
\begin{equation}
P =  - \frac{{\varepsilon _0 E}}{2}^2  - e_0 z\psi \left( {n_ +   - n_ -  } \right) + \mu _ +  n_ +   + \mu _ -  n_ -   + \left\langle {\mu _\omega  \left( \omega  \right)\rho \left( \omega  \right)} \right\rangle _\omega   + Ts
\label{eq:11}
\end{equation}
Substituting Eq. (\ref{eq:5}) in the above equation, we get the following equation 
\begin{equation}
\begin{array}{l}
 P =  - \frac{{\varepsilon _0 E}}{2}^2  - e_0 z\psi \left( {n_ +   - n_ -  } \right) + \left( {k_B T\ln \left( {n_{ + b} /N_b } \right) + V_ +  \lambda _b } \right)n_ +   + \left( {k_B T\ln \left( {n_{ - b} /N_b } \right) + V_ +  \lambda _b } \right)n_ -   \\ 
  + \left\langle {\left( {k_B T\ln \left( {\rho \left( \omega  \right)\Delta \Omega _i /N_b }\right) + V_w \lambda _b } \right)\rho _w \left( \omega  \right)} \right\rangle _\omega  \\ + k_B T\left[{N\ln N-\sum\limits_{i = \{+ , - \} }^{} {n_i \ln n_i }-\mathop {\lim }\limits_{k \to \infty } \sum\limits_{i = 1}^k {\left\{ {\left[ {\rho \left( {\omega _i } \right)\Delta \Omega _i } \right]\ln \left[ {\rho\left( {\omega _i } \right)\Delta \Omega _i } \right]}\right\}}}\right] \\ 
  = - \frac{{\varepsilon _0 E}}{2}^2 - e_0 z\psi \left( {n_+ - n_ - } \right) + k_B Tn_ +  \ln \left( {\frac{{n_{ + b} }}{{N_b }}\frac{N}{{n_ + }}} \right) + k_B Tn_ -  \ln \left( {\frac{{n_{ - b} }}{{N_b }}\frac{N}{{n_ - }}} \right) \\+ k_B T\left\langle {\rho \left( \omega  \right)\ln \left( {\frac{{\rho _b \left( \omega  \right)\Delta \Omega _i }}{{N_b }}\frac{{N_{} }}{{\rho \left( \omega  \right)\Delta \Omega _i }}} \right)} \right\rangle _\omega+ \lambda _b. 
 \end{array}
\label{eq:12}
\end{equation}
As the distance between the charged surfaces approaches the positive infinity,  $P=P_{bulk}$. 
In consequence, we obtain  $\lambda_b=P_{bulk}$ from Eq. (\ref{eq:12}).

We can find the formula for osmotic pressure by comparing  the above fact, Eq. (\ref{eq:12}) and Eq. (\ref{eq:13})
\begin{equation}
\begin{array}{l}
P_{osm} =  - \frac{{\varepsilon _0 E}}{2}^2  - e_0 z\psi \left( {n_ +   - n_ -  } \right) + k_B Tn_ +  \ln \left( {\frac{{n_{ + b} }}{{N_b }}\frac{N}{{n_ +  }}} \right) + k_B Tn_ -  \ln \left( {\frac{{n_{ - b} }}{{N_b }}\frac{N}{{n_ -  }}} \right)  \\ +
k_B T\left\langle {\rho \left( \omega \right)\ln \left( {\frac{{\rho _b \left( \omega  \right)\Delta \Omega _i }}{{N_b }}\frac{{N_{} }}{{\rho \left( \omega  \right)\Delta \Omega _i }}} \right)} \right\rangle _\omega.  
\end{array}
\label{eq:14}
\end{equation}
On the other hand, Eq. (\ref{subeq:4}) can be rewritten as follows.
\begin{equation}
n_{ + b} e^{ - V_ +  h - \beta ze\phi }  + n_{ - b} e^{ - V_ -  h + \beta ze\phi }  + n_{wb} e^{ - V_w h} \frac{{\sinh (p_0 \beta E)}}{{p_0 \beta E}} = n_{ + b}  + n_{ - b}  + n_{wb}  = N_b 
\label{eq:15}
\end{equation}

Rearranging Eq. (\ref{eq:15}) also results in a new relation
\begin{equation}
N = n_ +   + n_ -   + n_w  = \frac{{n_{ + b} \exp \left( { - V_ +  h - e_0 z\beta \psi } \right) + n_{ - b} \exp \left( { - V_ -  h + e_0 z\beta \psi } \right) + n_{wb} e^{ - V_w h} \frac{{\sinh (p_0 \beta E)}}{{p_0 \beta E}}}}{D} = \frac{{N_b }}{D}
\label{eq:16}
\end{equation}
Substituting the above relation in Eq. (\ref{eq:14}) and using the condition of volume conservation, we eventually obtain the following relation 
\begin{equation}
\begin{array}{l}
 P_{osm} =  - \frac{{\varepsilon _0 E}}{2}^2  - e_0 z\psi \left( {n_ +   - n_ -  } \right) + k_B Tn_ +  \ln \left( {\frac{1}{{\exp \left( { - V_ +  h - e_0 z\beta \psi } \right)}}} \right) + k_B Tn_ -  \ln \left( {\frac{1}{{\exp \left( { - V_ -  h + e_0 z\beta \psi } \right)}}} \right) +  \\ 
  + k_B T\left\langle {\rho \left( \omega  \right)\ln \left( {\frac{{\rho _b \left( \omega  \right)\Delta \Omega _i }}{{N_b }}\frac{N}{{\rho \left( \omega  \right)\Delta \Omega _i }}} \right)} \right\rangle _\omega   \\ 
  =  - \frac{{\varepsilon _0 E}}{2}^2  + k_B T(n_ +  V_ +   + n_ -  V_ -  )h +\\
+k_B T\left( {n_{wb} V_w h\exp \left( { - V_w h} \right)\frac{{\sinh \left( {p_0 \beta E} \right)}}{{p_0 \beta E}}/D + \exp \left( { - V_w h} \right)\left\langle {\exp (p_0 \beta E\cos \omega )(p_0 \beta E\cos \omega)} \right\rangle } \right) =  \\ 
  =  - \frac{{\varepsilon _0 E}}{2}^2  + k_B T(n_ +  V_ +   + n_ -  V_ -   + n_w V_w )h + k_B T\left( {\exp \left( { - V_w h} \right)\left\langle {\exp (p_0 \beta E\cos \omega )\left( {p_0 \beta E\cos \omega} \right)} \right\rangle _\omega} \right) \\ 
  =  - \frac{{\varepsilon _0 E}}{2}^2  + k_B Th + k_B Tn_{wb} \exp \left( { - V_w h} \right)/D\left\langle {\exp (p_0 \beta E\cos \omega )\left( {p_0 \beta E\cos \omega} \right)} \right\rangle _\omega   =  \\ 
  =  - \frac{{\varepsilon _0 E}}{2}^2  + k_B Th - k_B Tn_{wb} \exp \left( { - V_w h} \right)/D\left( {\cosh \left( {p_0 \beta E} \right) - \frac{{\sinh \left( {p_0 \beta E} \right)}}{{\left( {p_0 \beta E} \right)}}} \right) =  \\ 
  =  - \frac{{\varepsilon _0 E}}{2}^2  + k_B Th - k_B Tn_w \left( {\cosh \left( {p_0 \beta E} \right)\frac{{\left( {p_0 \beta E} \right)}}{{\sinh \left( {p_0 \beta E} \right)}} - 1} \right) =- \frac{{\varepsilon _0 E}}{2}^2  + k_B Th - k_B Tn_w \left({p_0 \beta E} \right)L\left({p_0 \beta E} \right).
 \end{array}
\label{eq:17}
\end{equation}
Eq. (\ref{eq:17}) corresponds to the osmotic pressure between two charged surfaces.
This osmotic pressure is constant across the channel, and therefore, can be estimated at any point
within the channel.
We can easily identify that the formula Eq. (\ref{eq:17}) is for a more general situation containing ones of Poisson-Boltzmann, Bikerman, Langevin-Poisson-Boltzmann, Modified Langevin-Poisson-Boltzmann, Langevin-Bikerman approach \cite%
{Chapman_PhilosMag_1913, Bikerman_PhilosMag_1942, Gongadze_GPB_2011, Velikonja_IJMS_2013, Das_JCP_2013, Gongadze_EA_2014} and the approach with ion size effect and without solvent polarization \cite%
{Boschitsch_JCC_2012}.
For the simplicity of calculation, we can express all the quantities in dimensionless forms as
\begin{equation}
\begin{array}{l}
 \bar x = x/\lambda, \bar \psi  = e_0 z\beta \psi ,\bar d = d/\lambda,
 \bar \varepsilon _r  = \varepsilon _r /\varepsilon _p, \lambda  = \sqrt {\frac{{\varepsilon _0 \varepsilon _p k_B T}}{{2n_b e_0^2 z^2 }}} ,\bar h = hV_w , \\ 
 \bar V = V/V_w ,\eta  = \frac{1}{{n_{wb} V_w }},\bar n_b  = \frac{{n_b }}{{n_{wb} }},p_0 \beta E = p_0 \sqrt {\frac{{2n_b }}{{\varepsilon _0 \varepsilon _p k_B T}}} \frac{{d\bar \psi }}{{d\bar r}} = \chi \bar E ,
\end{array}
\label{eq:18}
 \end{equation}
Based on the dimensionless parameters, Eq. \ref{eq:5} and Eq. \ref{eq:6} can be rewritten as follows
\begin{equation}
\bar n_{ + b} \left( {e^{ - \bar\psi  - \bar V_ + \bar h}  - 1} \right) +\bar n_{ - b} \left(  {e^{\bar \psi -\bar V_ - \bar h }  - 1} \right) +  \left( {e^{ -\bar  h} \frac{{\sinh (\chi \bar E)}}{{\chi \bar E}} - 1} \right) = 0,
\label{eq:20}
\end{equation}

\begin{equation}
\frac{d}{{d\bar x}}\left( {\bar \varepsilon _r \frac{{d\bar \psi }}{{d\bar x}}} \right) = \frac{\eta }{2}\frac{{\exp \left( {\bar \psi  - \bar V_ -  \bar h} \right) - \exp \left( { - \bar \psi  - \bar V_ +  \bar h} \right)}}{{\frac{{\sinh \left( {\chi \bar E} \right)}}{{\chi \bar E}}\exp \left( { - \bar h} \right) + \bar n_b \bar V_ -  \exp \left( {\bar \psi  - \bar V_ -  \bar h} \right) + \bar n_b \bar V_ +  \exp \left( { - \bar \psi  - \bar V_ +  \bar h} \right)}}.
\label{eq:19}
 \end{equation}
\section{Results and Discussion}
All the calculations in the present study are performed by using the fourth order Runge-Kutta method combined with shooting method. For clarity, we choose 0.01M for the ionic concentration in the buk electrolyte solution and  298K for the temperature.
\subsection{Similarly Charged Surfaces}
Without loss of generality, we assume that the surfaces are positively charged. 

\begin{figure}
\begin{center}
\includegraphics[width=1\textwidth]{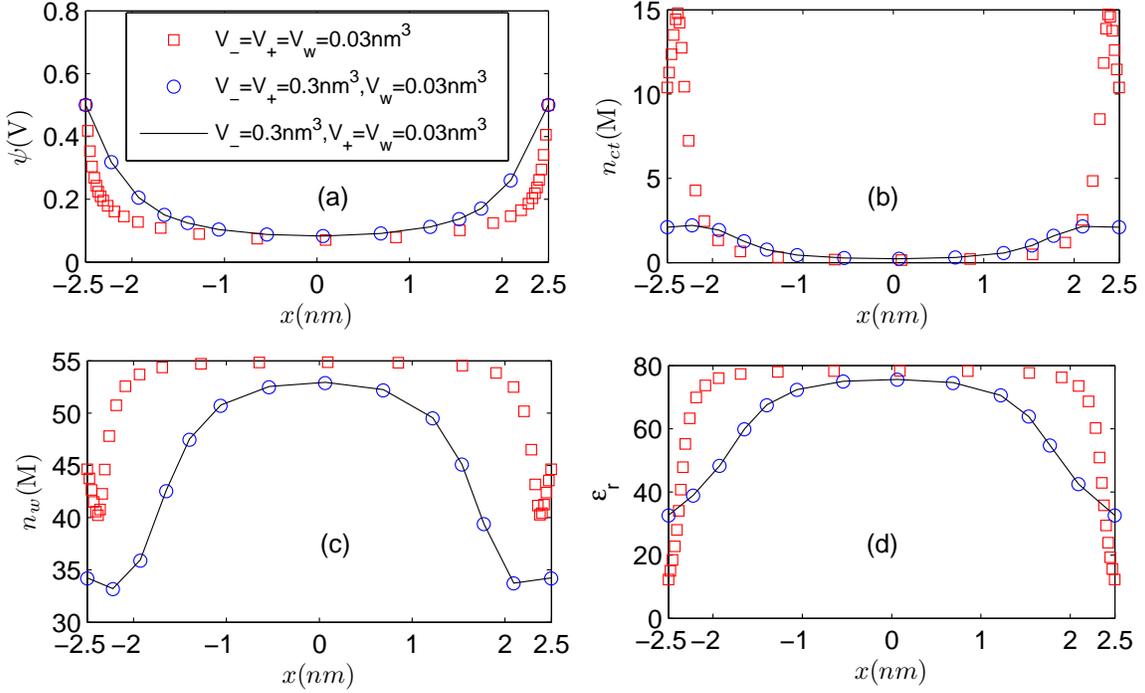}
\caption{(Color online)For similarly charged surfaces, variation of electrostatic potential (a), the number density of counterions (b), the number density and water molecules(c) and  the permittivity (d) with the position for different sets of ion sizes. The separation distance between charged surfaces is  $H=5$nm and the surface potential is $\psi\left(x=H/2)\right)=\psi\left(x=-H/2)\right)=+0.5$V.
}
\label{fig:1}
\end{center}
\end{figure}
\begin{figure}
\begin{center}
\includegraphics[width=0.9\textwidth]{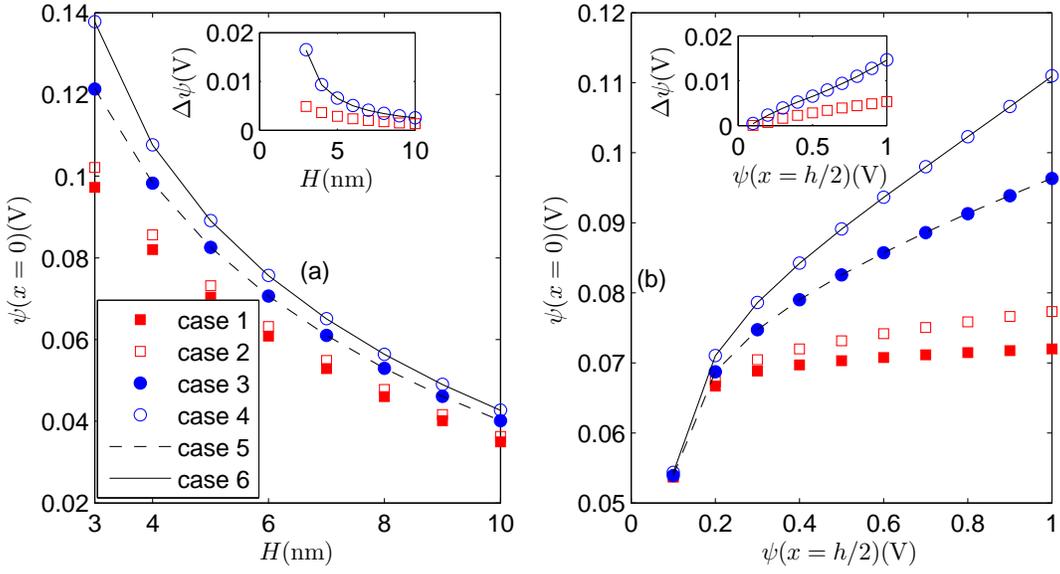}
\caption{(Color online) For similarly charged surfaces, (a) Variation of the centerline potential  with the separation distance between the charged surfaces for $\psi\left(x=H/2)\right)=\psi\left(x=-H/2)\right)=+0.5$V. (b) Variation of the centerline potential with the surface potential for different sets of ion sizes. The separation distance between charged surfaces is  $H=5$nm.}
\label{fig:2}
\end{center}
\end{figure}
\begin{figure}
\begin{center}
\includegraphics[width=0.9\textwidth]{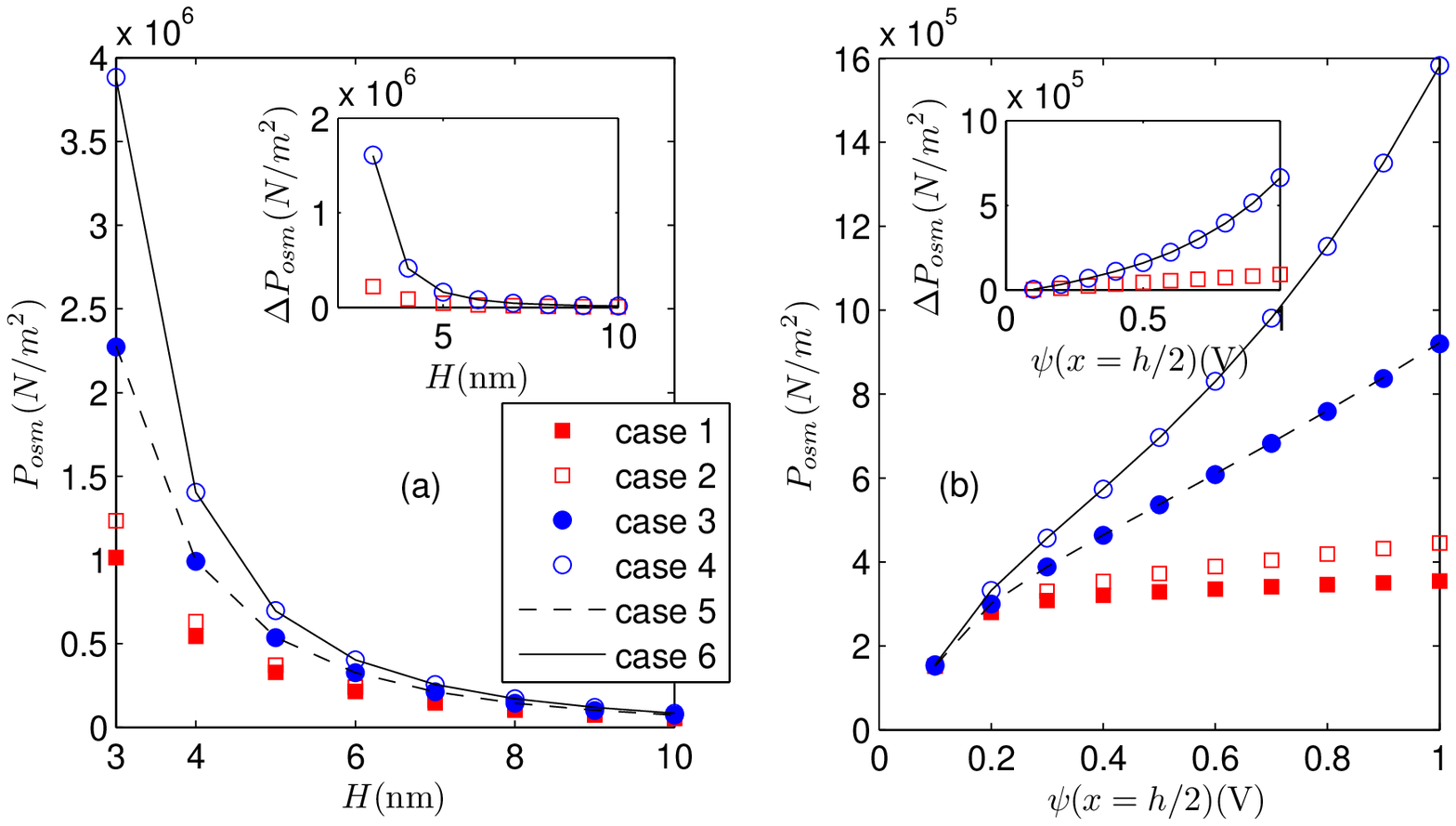}
\caption{(Color online) For similarly charged surfaces, (a) Variation of osmotic pressure with the separation distance between the charged surfaces for $\psi\left(x=H/2)\right)=\psi\left(x=-H/2)\right)=+0.5$V. (b) Variation of the osmotic pressure as a function of the surface potential for different sets of ion sizes and the separation distance between charged surfaces, $H=5$nm.}
\label{fig:3}
\end{center}
\end{figure}
Fig. \ref{fig:1}(a) shows the electrostatic potential profile between similarly charged surfaces for the case where the distance between charged surfaces is $H=5$nm  and the surface potential is $\psi\left(x=h/2\right)=0.5$V. The cases for $\left(V_-=V_+=V_w=0.03nm^3\right)$, $\left(V_-=V_+=0.3nm^3, V_w=0.03nm^3\right)$ and $\left(V_-=0.3nm^3, V_+=V_w=0.03nm^3\right)$   are represented by squares, circles and solid line, respectively. In Fig. 1(a), it is illustrated that the electrostatic potential has a symmetric distribution attributed to the geometry of this system and the sign of the surface potentials. We can also find that an increase in counterion size makes the electrostatic potential increase in the region between two charged surfaces. This is understood by the fact that the screening property of a counterion with a large size is weaker than corresponding one of a counterion with a smaller size. Importantly, we should emphasize that the electrostatic potential profile is not related to coion size, as in \cite%
{Sin_EA_2016, Sin_CSA_2017}.

Fig. \ref{fig:1}(b) demonstrates the spatial dependence of the number density of counterions. For the cases when a counterion has the small size ($V_-=0.03nm^3$), the number density of counterions is larger than corresponding one for a counterion of the larger size ($V_-=0.3nm^3$)in the vicinity of a charged surface. Such a phenomenon is attributed to the excluded volume effect, as explained in \cite%
{Iglic_Bioelechem_2010}.

Fig. \ref{fig:1}(c) shows the transverse variation of the number density of water molecules. It is clearly seen that the density for a counterion of the large size is smaller than corresponding one for a counterion of the smaller size. 
This is explained by combining excluded volume effect of counterions and the fact that due to overlap of the electric double layers of similarly charged surfaces, the number density of counterions gets larger than the bulk value.

Fig.  \ref{fig:1}(d) shows the variation of the permittivity with the position between two charged surfaces. According to the permittivity formula of Eq. (\ref{eq:7}), the permittivity is strongly affected by the number density of water molecules and the electric field strength. Near the centerline between the charged surfaces, the permittivity is proportional to the number density of water molecules, since due to the geometrical symmetry of the present system the electric field strength is zero at the centerline between two charged surfaces. However, in the vicinity of a charged surface, in spite of the fact that the number density of water molecules for the case when a counterion has the large size is smaller than corresponding one for a counterion of the smaller size, the permittivity for the former case is larger than one for the latter case. That is why a high electric field strength lowers the permittivity \cite%
{Booth_JCP_1955} .

Fig.  \ref{fig:2}(a) shows the electrostatic potential at the centerline between two charged surfaces as a function of the  separation distance between the surfaces.
In Figs. (\ref{eq:2}, \ref{eq:3}, \ref{eq:5}, \ref{eq:6}), we use an identical convention for the studied cases.
Case 1, Case 3 and Case 5 mean  $\left(V_-=V_+=V_w=0.03nm^3\right)$, $\left(V_-=V_+=0.3nm^3, V_w=0.03nm^3\right)$ and  $\left(V_-=0.3nm^3, V_+=V_w=0.03nm^3\right)$ for the case when both solvent polarization and ion sizes are considered, respectively.
Case 2 , Case 4 and Case 6 represent   $\left(V_-=V_+=V_w=0.03nm^3\right)$, $\left(V_-=V_+=0.3nm^3, V_w=0.03nm^3\right)$ and  $\left(V_-=0.3nm^3, V_+=V_w=0.03nm^3\right)$ for the case accounting only for ion sizes, respectively.

For similarly charged surfaces, the electrostatic potential at the centerline between them is a main characteristics that illustrates overlap of electric double layers.
 It is noticeable that under the boundary condition of a given potential, solvent polarization lowers the potential at the centerline. In fact, as pointed out in \cite%
{Das_PRE_2011}, ion size effect enhances electrostatic potential at the centerline due to lowering of screening property of ions. However, solvent polarization lowers the permittivity as shown in Fig. 1(d). From the viewpoint of physics, lowering permittivity yields enhancement of magnitude of electric field strength in an electrolyte.  Since at any position between two charged surfaces the permittivity value is not higher than the bulk value of the permittivity, the electric field strength is not lower than corresponding one for the case with only ion size effects. Finally for the case with solvent polarization, the centerline potential is lower than one for the case without the effect.
As can be expected, for either case, an increase in the separation distance between two charged surfaces decreases the centerline potential. This is deduced by using the fact that the longer the distance between two charged surfaces, the weaker the overlap of electric double layers of the charged surfaces. 
The inset in Fig. 2(a) shows that $\Delta \psi$, the difference in the centerline potential between the cases with and without solvent polarization, decreases with the distance between charged surfaces. It can be explained by the fact that as the distance between charged surfaces gets longer, the centerline potential for either case tends to zero.

 Fig.  \ref{fig:2}(b) shows the dependence of the centerline potential on surface potential at charged surfaces.
It is clearly demonstrated that an increase in the surface potential allows the centerline potential to increase. This is attributed to the fact that for a given distance between two charged surfaces, an increase in the surface potential enhances the overlap of electric double layers of two charged surface and therefore induces the increase in the centerline potential.The inset in Fig. 2(b) shows that the difference in the centerline potential between the cases with and without solvent polarization, increases with increasing the surfaces potential. It can be explained by the fact that an increase in the surface potential involves enhancement of solvent polarization .

Fig. \ref{fig:3}(a) shows the variation of osmotic pressure with the distance separation between two charged surfaces. As one can see, a large ion size makes repulsive osmotic pressure higher than for the smaller one and solvent polarization decreases the osmotic pressure. For all the cases, an increase in the distance between charged surfaces yields diminished osmotic pressure. The above facts are attributed to the following fact.
As shown in \cite%
{Sin_EA_2016}, the value of h increases with increasing the electric potential. On the other hand, as mentioned above, ion size effects increases the centerline potential. Namely, h increases with ion size. At the centerline, the electric field strength is zero due to the geometrical symmetry of this problem. As a result, the formula for osmotic pressure between the similarly charged surfaces is rewritten as follows:
\begin{equation}
P_{osm}  = k_B Th(x = 0).
\label{eq:22}
\end{equation}
Consequently, the osmotic pressure increases with counterion size, decreases when we consider solvent polarization. As the distance between charged surfaces approaches infinity, it also tends to zero.
The inset of Fig. \ref{fig:3}(a) shows that an increase in the distance between the surfaces allows $\Delta P_{osm}$,  being the difference in osmotic pressure between the cases with and without solvent polarization, to be decreased. Combining the Eq. (\ref{eq:22}) and the explanation for Fig. \ref{fig:2}(a) provides the reason for the phenomenon. 

Fig. \ref{fig:3}(b) shows the surface potential dependence of osmotic pressure between two charged surfaces. Fig. \ref{fig:3}(b) demonstrates that an increase in the surface potential induces an increase of osmotic pressure. This phenomenon is elucidated by combining Eq. (\ref{eq:22}) and the fact that the centerline potential increases with increasing the surface potential.
Although we have treated the cases of positively charged surfaces, the same is true for the cases of negatively charged surfaces.

\begin{figure}
\begin{center}
\includegraphics[width=0.5\textwidth]{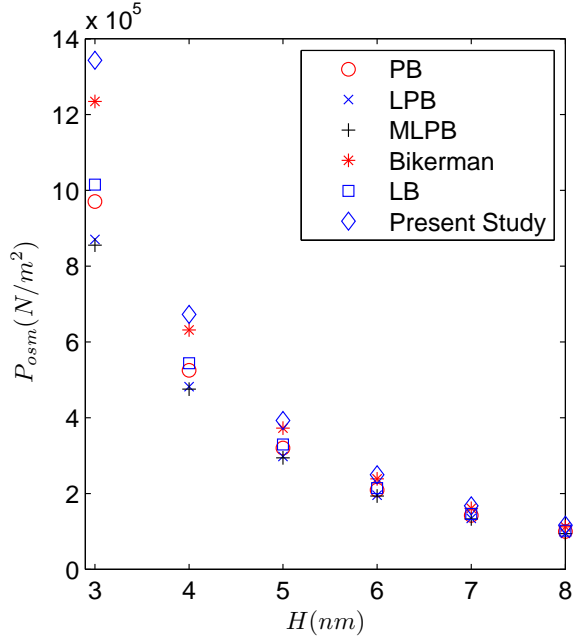}
\caption{(Color online)  For similarly charged surfaces, Variation of osmotic pressure with the separation distance between the charged surfaces for $\psi\left(x=H/2)\right)=\psi\left(x=-H/2)\right)=+0.5$V for different types of electric double layer model.  Circles, crosses, plus signs, asterisks, squares and diamonds stand for Poisson-Boltzmann(PB), Langevin-Poisson-Boltzmann(LPB), Modified Langevin-Poisson-Boltzmann(MLPB), Bikerman, Langevin-Bikerman(LB) and present approach($V_-=V_+=0.1nm^3$), respectively.}
\label{fig:7}
\end{center}
\end{figure}

Fig.\ref{fig:7} shows the dependence of osmotic pressure on the distance separation between the surfaces for similarly charged surfaces ($\psi\left(x=H/2)\right)=\psi\left(x=-H/2)\right)=+0.5$V) for Poisson-Boltzmann(PB), Langevin-Poisson-Boltzmann(LPB), Modified Langevin-Poisson-Boltzmann(MLPB), Bikerman, Langevin-Bikerman(LB) and the present approach($V_-=V_+=0.1nm^3$).  Here circles, crosses, plus signs, asterisks, squares and diamonds stand for Poisson-Boltzmann, Langevin-Poisson-Boltzmann, Modified Langevin-Poisson-Boltzmann, Bikerman, Langevin-Bikerman and present approach($V_-=V_+=0.1nm^3$), respectively. Comparison of the osmotic pressure calculated using different types of electric double layer model shows the following facts. 

First, we confirm again that the models with solvent polarization(Langevin-Poisson-Boltzmann and Langevin-Bikerman) predict lower values of osmotic pressure than for corresponding models (Poisson-Boltzmann and Bikerman) without solvent polarization, respectively.   

It is also identified that the models with larger volumes of ions result in larger values of osmotic pressure. In other words, Poisson-Boltzmann approach(point-like ion) predicts lower values of osmotic pressure than for Bikerman approach($V_-=V_+=V_w=0.03nm^3$). And Langevin-Poisson-Boltzmann approach(point-like ions) yields lower values of osmotic pressure than for Langevin-Bikerman approach($V_-=V_+=V_w=0.03nm^3$) which provides lower osmotic pressures than for present approach($V_-=V_+=0.1nm^3$). 

Finally, it should be pointed out that  modified Langevin-Poisson-Boltzmann approach predicts slightly lower values of osmotic pressure than for Langevin-Poisson-Boltzmann approach. The difference is attributed to a stronger decrease in permittivity due to consideration of cavity field as pointed out in \cite%
{Iglic_Bioelechem_2012}. However, since the difference in osmotic pressure for the two approaches is small, we confirm that the present approach without consideration of cavity field is quite reliable for computing osmotic pressure.

\subsection{Oppositely Charged Surfaces}

\begin{figure}
\begin{center}
\includegraphics[width=0.9\textwidth]{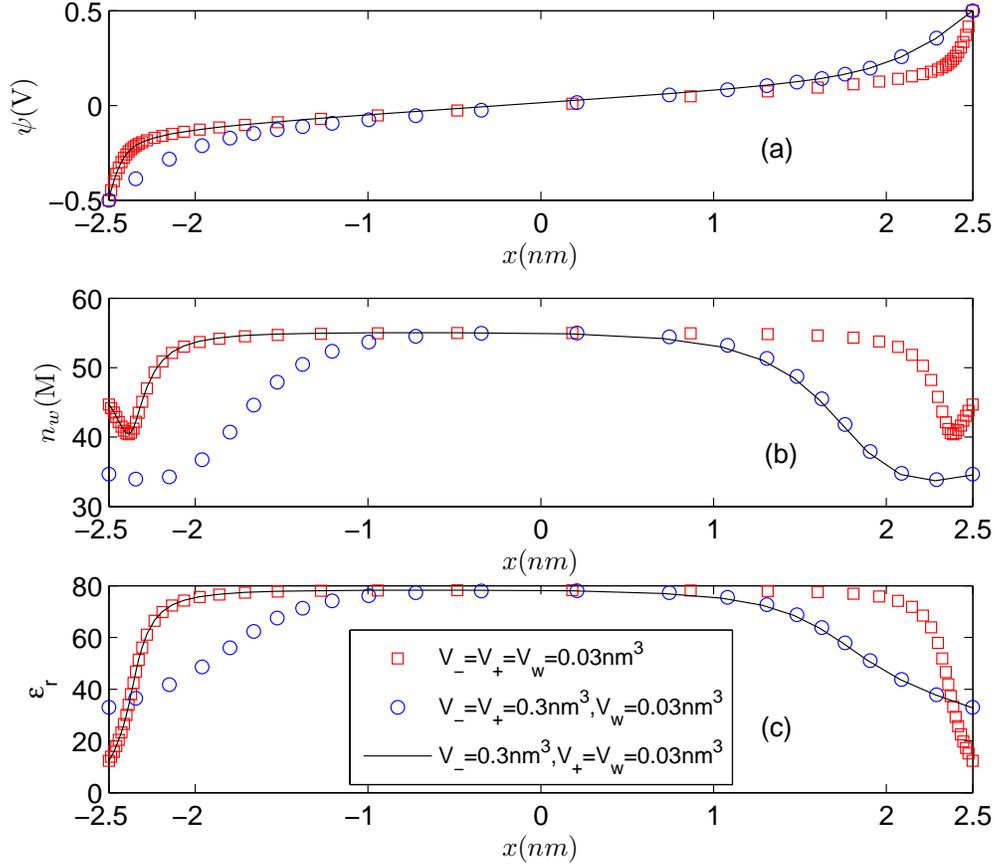}
\caption{(Color online)For oppositely charged surfaces, variation of electrostatic potential (a), the number density and water molecules(b) and  the permittivity (c) with the position for different sets of ion sizes. The separation distance between charged surfaces is  $H=5$nm and the surface potential is $ \psi\left(x=H/2)\right)=-\psi\left(x=-H/2)\right)=+0.5$V}
\label{fig:4}
\end{center}
\end{figure}

Fig. \ref{fig:4}(a) shows the electrostatic potential profile for oppositely charged surfaces. The symbols in Fig. (\ref{fig:4}) have the same meanings as in Fig. (\ref{fig:1}). As can be seen, for the cases when negative and positive ions are the same size, the profile has point symmetry in the position. This is attributed to the geometrical and electrical symmetry of the system. For the case of unequal sizes of ions($V_-=0.3nm^3, V_+=V_w=0.03nm^3$), the profile has not point symmetry. For the case of unequal sizes, near the surface with negative potential the profile is equal to one for the case of $V_-= V_+=V_w=0.03nm^3$, near the surface with positive potential the profile is equivalent to one for the case of $V_-= V_+=0.3nm^3, V_w=0.03nm^3$. As mentioned-above, this fact is easily understood by the fact that electric double layer near a charged surface is determined mainly by counterions.

Fig. \ref{fig:4}(b) and Fig. \ref{fig:4}(c) show the number density of water molecules and the permittivity between the oppositely charged surfaces, respectively. 
Those quantities exhibits such a behavior as in Fig. \ref{fig:4}(a) due to the same reason. 

\begin{figure}
\begin{center}
\includegraphics[width=0.9\textwidth]{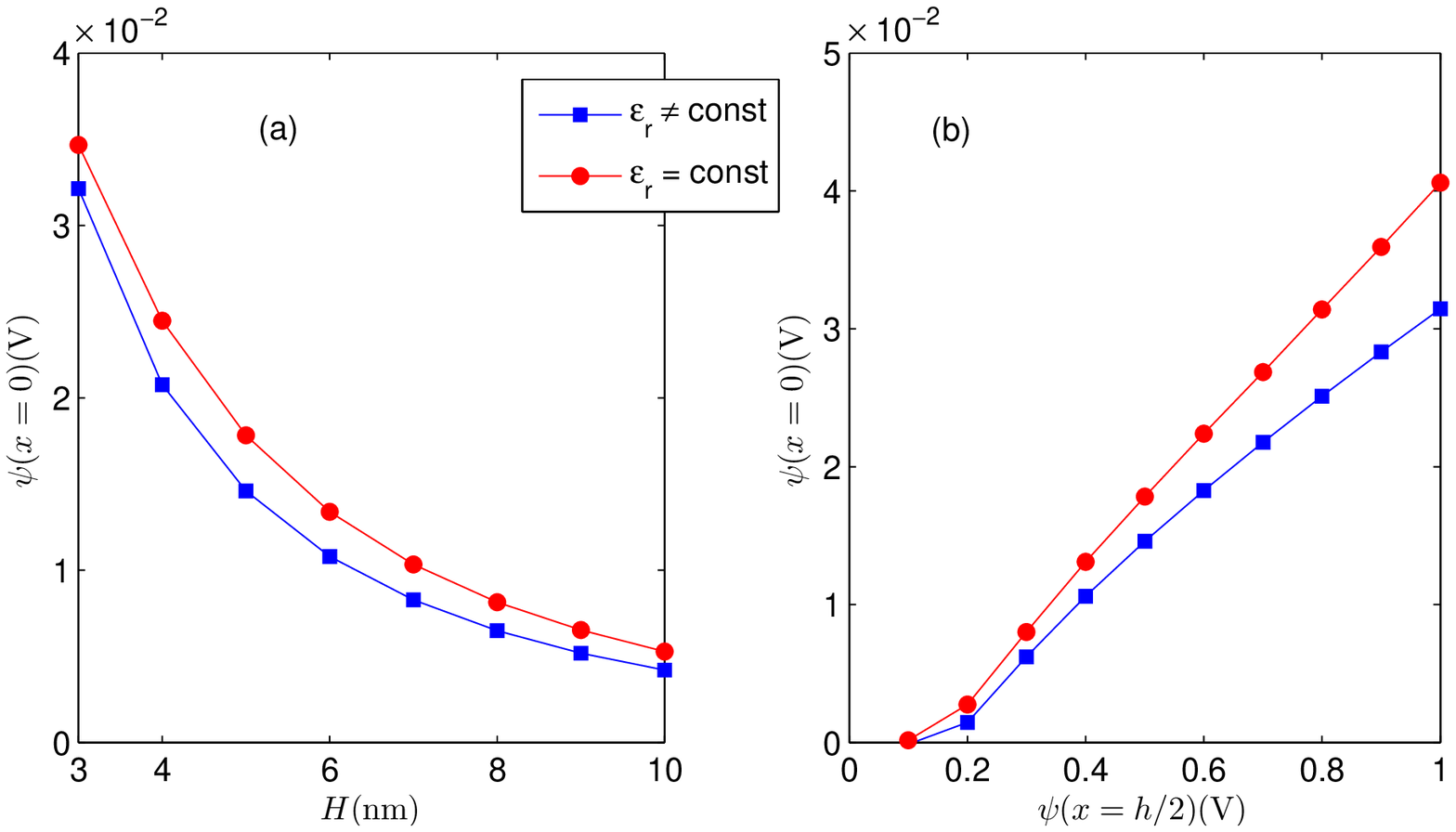}
\caption{(Color online)  For oppositely charged surfaces, (a) Variation of the centerline potential with the separation distance between the charged surfaces for $\psi\left(x=H/2)\right)=-\psi\left(x=-H/2)\right)=+0.5$V. (b) Variation of the centerline potential with the surface potential for the case of unequal ion sizes($V_-=0.3nm^3, V_+=V_w=0.03nm^3$) and the separation distance between charged surfaces, $H=5$nm.}
\label{fig:5}
\end{center}
\end{figure}

Fig. \ref{fig:5}(a) show the centerline potential according to the distance between charged surfaces for the case of $V_-=0.3nm^3, V_+=V_w=0.03nm^3$.

\begin{figure}
\begin{center}
\includegraphics[width=0.9\textwidth]{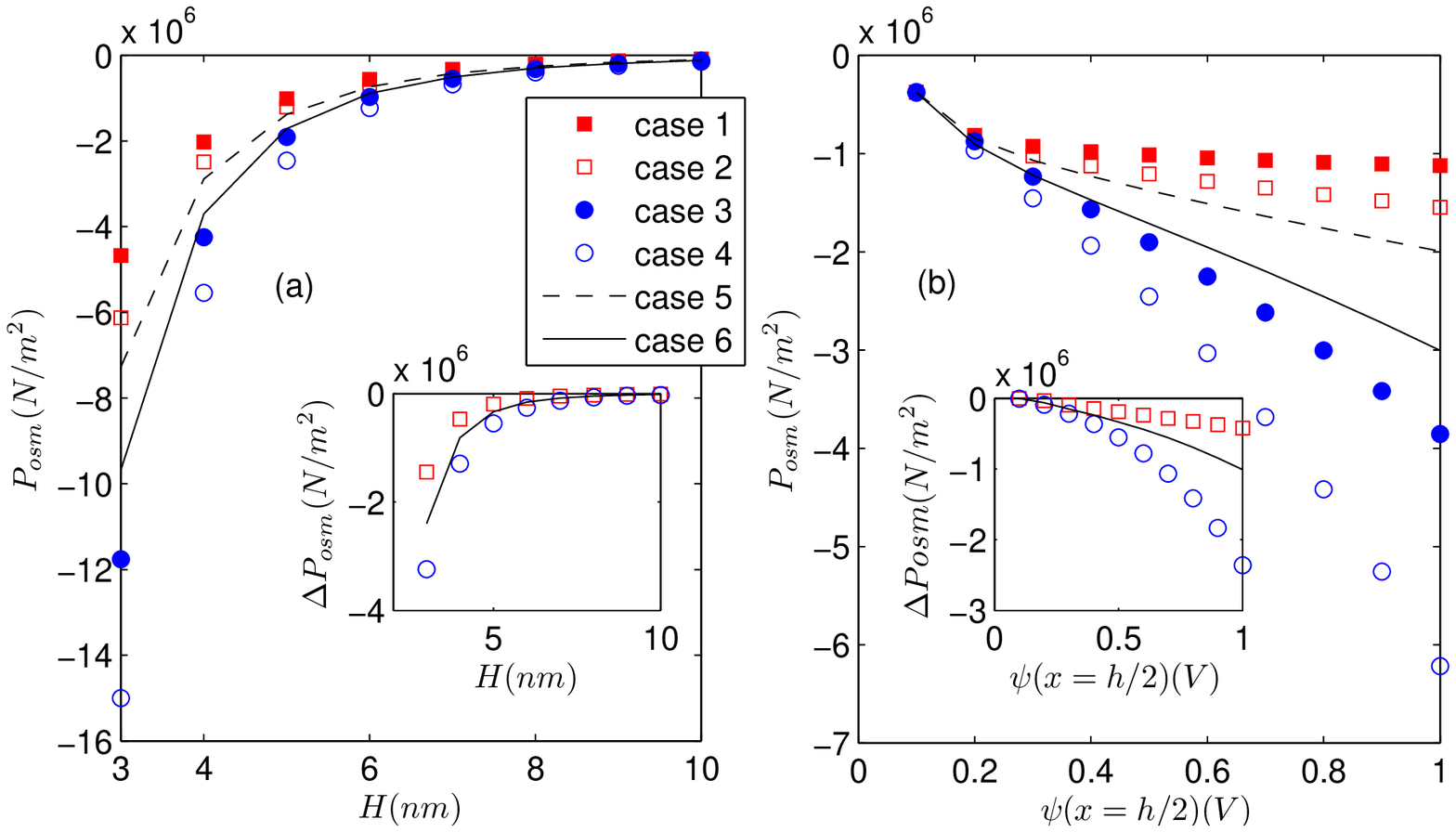}
\caption{(Color online)  For oppositely charged surfaces, (a) Variation of osmotic pressure with the separation distance between the charged surfaces for $\psi\left(x=H/2)\right)=-\psi\left(x=-H/2)\right)=+0.5$V. (b) Variation of the osmotic pressure with the surface potential for different sets of ion sizes and the separation distance between charged surfaces, $H=5$nm.}
\label{fig:6}
\end{center}
\end{figure}

Due to the symmetry, for different cases the centerline potential is zero at any distance and any magnitude of surface potential.  It is clearly seen that for the cases with and without solvent polarization, the centerline potential decreases with the distance between two charged surfaces. The value for the case with solvent polarization is lower than that for the case without solvent polarization. The fact is explained by the fact that the low permittivity induces high magnitude of electric field strength.    

Fig. \ref{fig:5}(b) shows the centerline potential with the surface potential. Due to the same reason in Fig. 5(a), the centerline potential increases with increasing magnitude of the surface potential. It is noticeable that the difference in the centerline potential between the two charged surfaces is enhanced with the surface potential.   

Fig. \ref{fig:6}(a) is a graph of the attractive osmotic pressure as a function of the distance between charged surfaces. As one can see, an increase in ion size involves the enhancement of the attractive osmotic pressure due to the same reason in the Fig. 3(a). Unlike the case of similarly charged surfaces, both sizes of a positive and negative ion are important for determining the pressure. This is understood by considering the fact that counter-ions of the two surfaces have the different sign of electric charge. 
The inset of Fig. \ref{fig:6}(a) shows that as the distance between the oppositely charged surfaces increases, the attractive osmotic pressure vanishes due to the same reason as in the inset of Fig. \ref{fig:3}(a)

Fig. \ref{fig:6}(b) shows the variation of osmotic pressure with the surface voltage.
Fig. \ref{fig:6}(b) demonstrates that as the surface potentials increases, the attractive osmotic pressure increases due to the same reason as in the inset of Fig. \ref{fig:3}(b).
The inset of Fig. \ref{fig:6}(b) shows that as the surface potentials increases, the difference in the attractive osmotic pressures between charged surfaces for the cases with and without solvent polarization, increases due to the same reason as in the inset of Fig. \ref{fig:3}(b).

Summarizing Figs. (\ref{fig:2}, \ref{fig:3}, \ref{fig:6}), it is elucidated that for the case of constant surface potential, solvent polarization reduces ion size effect. Namely, the centerline potential and osmotic pressure between two charged surfaces are diminished for the case with solvent polarization than for the case without the effect. This result distinguishes our theory from one of \cite%
{Das_JCP_2013}. In fact, they considered uniform size effect  as well as solvent polarization, while they asserted that experiment results are understood by the increase due to only the consideration of solvent polarization.
We believe that the reason is not so simple as their ones.  On one hand, ion size effect, as shown in Figs. (\ref{fig:2}, \ref{fig:3}, \ref{fig:5}, \ref{fig:6}), enhances the centerline potential and osmotic pressure for similar or oppositely charged surfaces. 
On the other hand, the consideration of solvent polarization involves lowering the properties. As a result,  solvent polarization reasonably lowers excessive increases in centerline potential and osmotic pressure due to ion size effect.  We conclude that simultaneous considerations of solvent polarization and ion size effect is mandatory for elucidating experimental results. 

Also, the present theory can take into account the difference in size not only between ions but also between ions and a water molecule. 
This fact demonstrates the clear advantage that unlike in the previous theory \cite%
{Das_JCP_2013} where ions and water molecules have the same size, the present method can treat more realistic situations where sizes of ions and water molecules are not equal to each other. 

Our results can be compared with Monte Carlo simulation describing orientational ordering of solvent dipoles and ion size effects. The simulation will require vast computational cost due to a large number of degrees of freedom for the present study.   

\section{Conclusions}
Using a mean-field theory accounting for solvent polarization and unequal size effect, we have studied electrostatic properties between two charged surfaces.
We have shown that the electrostatic properties are unconditionally symmetrical about the centerline between similarly charged surfaces but not between oppositely charged surfaces.

We have demonstrated that for the case of similarly charged surfaces, electrostatic properties are determined mainly by counterions but not by coions. In contrast to the case,  the properties for oppositely charged surfaces are determined by both positive and negative ions.  
Moreover, the centerline potential, being a quantity representing overlap of electric double layers of similarly charged surfaces, and the osmotic pressure between the surfaces  increase with the counterion size. 
Most importantly, we have found that under the condition of constant surface potential, the consideration of solvent polarization reduces the centerline potential and osmotic pressure augmented  by ion size effect.    

\nocite{*}
\bibliography{aipsamp}

\end{document}